\pdfoutput=1
 \documentclass[aps,prl,reprint,twocolumn,showpacs,superscriptaddress,%
nobibnotes,nofootinbib]{revtex4-1}  

\usepackage{tabularx}
\makeatletter
\def\hlinewd#1{%
\noalign{\ifnum0=`}\fi\hrule \@height #1 %
\futurelet\reserved@a\@xhline}
\makeatother
 \usepackage{auncial}
\usepackage{scrextend}
\usepackage[T1]{fontenc}
\usepackage{relsize}
\usepackage{amsmath}
\usepackage{amssymb}
\usepackage{epsfig}
\usepackage{graphicx}
\usepackage{hyperref}
\usepackage{dcolumn}   
\usepackage{tabu}
\usepackage{boldline}
\usepackage{slashed}
\usepackage{multirow}
\usepackage{color}
\usepackage[normal]{subfigure}
\usepackage{rotating}
\usepackage[margin=0.9in,a4paper]{geometry}
\usepackage[table]{xcolor}
\usepackage{enumitem}
\usepackage[utf8]{inputenc}
\usepackage{colortbl}
\usepackage{array,multirow}
\definecolor{nicered}{rgb}{0.7,0.1,0.1}
\definecolor{nicegreen}{rgb}{0.1,0.5,0.1}
\definecolor{red}{rgb}{1.0, 0, 0}

\def\eqs#1#2{{Eqs.~(\ref{#1})--(\ref{#2})}}

\def\fig#1{{Fig.~\ref{#1}}}

\def\abs#1{\left| #1\right|}

\def\gsim{\raise0.3ex\hbox{$\;>$\kern-0.75em\raise-1.1ex\hbox{$\sim\;$}}}
\def\lsim{\raise0.3ex\hbox{$\;<$\kern-0.75em\raise-1.1ex\hbox{$\sim\;$}}}

\def\mb[#1]{\mathbf{#1}}

\renewcommand{\bar}{\overline}

\definecolor{LightCyan}{rgb}{0.88,1,1}
\definecolor{piggypink}{rgb}{0.99, 0.87, 0.9}
\definecolor{applegreen}{rgb}{0.55, 0.71, 0.0}
\definecolor{darkpastelgreen}{rgb}{0.01, 0.75, 0.24}
\definecolor{green-yellow}{rgb}{0.68, 1.0, 0.18}

\newcommand{\beq}{\begin{equation}}
\newcommand{\eeq}{\end{equation}}
\newcommand{\beqa}{\begin{eqnarray}}
\newcommand{\eeqa}{\end{eqnarray}}

\newcommand{\eV}{{\, \rm eV}}

\def\KSVZ{{\sf KSVZ}}
\def\DFSZ{{\sf DFSZ}}
\def\DFSZI{{\sf DFSZ-I}}

\def\DFSZI-II{{\sf DFSZ-I,II}}
\def\MI{{\sf M1}}
\def\MI{{\sf M1}}
\def\MII{{\sf M2}}
\def\MIII{{\sf M3}}
\def\MIV{{\sf M4}}

\newcommand{\Sec}[1]{ \medskip \noindent {\sl \bfseries #1}}

\newcommand{\eqn}[1]{Eq.~(\ref{#1})}
\newcommand{\eqns}[2]{Eqs.~(\ref{#1}),\,(\ref{#2})}
\newcommand{\Eqn}[1]{Eq.~(\ref{#1})}

\begin{document}



\title{Astrophobic Axions}

\author{Luca Di Luzio}
\affiliation{\normalsize \it 
Institute for Particle Physics Phenomenology, Department of Physics, Durham University, DH1 3LE, Durham, United Kingdom}
\author{Federico Mescia}
\affiliation{\normalsize \it 
Dept.~de F\'{\i}sica Qu\`antica i Astrof\'{\i}sica, Institut de Ci\`encies del Cosmos (ICCUB), 
Universitat de Barcelona, Mart\'i Franqu\`es 1, E08028 Barcelona, Spain
}
\author{Enrico Nardi}
\affiliation{\normalsize \it 
INFN, Laboratori Nazionali di Frascati, C.P.~13, 100044 Frascati, Italy
}
\author{Paolo Panci}
\affiliation{\normalsize \it 
Theoretical Physics Department, CERN, Geneva, Switzerland
}
\author{Robert Ziegler}
\affiliation{\normalsize \it 
Theoretical Physics Department, CERN, Geneva, Switzerland
}

\begin{abstract}
\noindent
We propose a class of axion 
models with generation dependent Peccei-Quinn charges for the known
fermions that allow to suppress the axion couplings to nucleons and
electrons. Astrophysical limits are thus relaxed, allowing for axion
masses up to ${\cal O}(0.1) \eV$. The axion-photon coupling remains
instead sizeable, so that next generation helioscopes will be able to
probe this scenario.  Astrophobia unavoidably implies flavor violating
axion couplings, so that experimental limits on flavor-violating
processes can provide complementary probes.  The astrophobic axion can
be a viable dark matter candidate in the heavy mass window, and can
also account for anomalous energy loss in stars.
\end{abstract}
%

\maketitle

\vspace{-20mm}
\Sec{Introduction.} 
%
%
One of the main mysteries of the standard model
(SM) is the absence of CP violation in strong interactions.  The most
elegant solution is provided by the Peccei-Quinn (PQ)
mechanism~\cite{Peccei:1977ur,Peccei:1977hh} which predicts the axion
as a low-energy remnant~\cite{Weinberg:1977ma,Wilczek:1977pj}.  The
axion is required to be extremely light and decoupled, and in a
certain mass range it can provide a viable dark matter (DM) candidate.
The Kim-Shifman-Vainshtein-Zakharov
(\KSVZ)~\cite{Kim:1979if,Shifman:1979if} and
Dine-Fischler-Srednicki-Zhitnitsky
(\DFSZ)~\cite{Zhitnitsky:1980tq,Dine:1981rt} axion models are
frequently used as benchmarks to assess experimental sensitivities and
to derive astrophysical bounds. However, constraining axion
properties solely on the basis of standard benchmarks can be too
restrictive, and exploring alternative models whose properties can
sizably deviate from those of \KSVZ\ and \DFSZ\ is highly desirable.
While it is conceptually easy to build models with suppressed axion-electron
couplings $g_{ae}$ \cite{Kim:1979if,Shifman:1979if,Kim:1984pt} or
axion-photon couplings $g_{a\gamma}$ \cite{Kaplan:1985dv,DiLuzio:2016sbl,DiLuzio:2017pfr}, it is
generally believed that a robust prediction of all axion models is an
unsuppressed axion-nucleon coupling $g_{aN}$. 
This is particularly important, because $g_{aN}$ is responsible for the often quoted
bound on the axion mass $m_a \lesssim 20 $ meV from the neutrino
burst duration of the SN1987A~\cite{Olive:2016xmw,Raffelt:2006cw}.
In this Letter we argue that a strong suppression of $g_{aN}$ is
instead possible in a class of \DFSZ-like models with
generation-dependent PQ charges.  Additional strong bounds on $m_a$
are obtained if, as in standard \DFSZ, $g_{ae}$ is unsuppressed, since
this can affect white-dwarf (WD) cooling rates and red giants (RG)
evolution in globular clusters~\cite{Raffelt:2006cw}.  However, a
suppression of $g_{ae}$ can be also arranged in our scenario.  Thus,
nucleophobia allows to relax the SN bound and electrophobia allows to
evade the WD/RG constraints, rendering viable masses up to
$m_a \sim 0.2$ eV. We denote such an axion as \emph{astrophobic},
although $g_{a\gamma}$ remains generically sizable, and could still
affect the evolution of horizontal branch (HB) stars.  Astrophobic
axions are interesting in many respects: $i)$ they render viable a
parameter space region well beyond the standard \DFSZ\ and \KSVZ\
benchmarks, yet still within the reach of the planned IAXO
helioscope~\cite{Irastorza:1567109}.
$ii)$~Nucleophobia necessarily implies flavor-violating (FV) axion
couplings to the quarks, so that
complementary searches can be carried out in flavor experiments.
$iii)$ Astrophobic axions can be non-standard DM in the heavy mass
window~\cite{Kawasaki:2014sqa,Ringwald:2015dsf,Co:2017mop} and $iv)$
can provide an explanation for various hints of anomalous energy loss
in stars~\cite{Giannotti:2015kwo,Giannotti:2017hny}.

\Sec{Axion coupling to nucleons.}
Let us first recall why $g_{aN}$ cannot be suppressed in \KSVZ\ and
\DFSZ\ models.  The relevant terms for this discussion are:
\begin{align} 
\label{LabelowEW}
\nonumber
& \mathcal{L}_{a} \supset 
\frac{\alpha_{\rm s}}{8 \pi} \frac{a}{f_a} G^a_{\mu\nu} \tilde G^{a,\mu\nu} 
 + \frac{\alpha }{8 \pi}\frac{E}{N}\frac{a}{f_a} F_{\mu\nu} \tilde F^{\mu\nu}  \\
&\!\!\!\! + \frac{\partial_\mu a}{2f_a}\!\! 
 \sum_{Q=U,D}\! \left[\bar Q_L \frac{X_{Q_L}}{N} \gamma^\mu Q_L +  
\bar Q_R  \frac{X_{Q_R}}{N} \gamma^\mu Q_R \right], 
\end{align}
where $N(E)$ are the QCD (QED) anomaly coefficients, $f_a= v_a/(2N)$ with
$v_a=\sqrt{2}\langle \phi \rangle$ the vacuum expectation value (VEV)
of the PQ symmetry breaking singlet field,
$\tilde G^{a,\mu\nu} = \frac{1}{2} \epsilon^{\mu\nu\rho\sigma}
G^{a}_{\rho\sigma}$,
$\tilde F^{\mu\nu} = \frac{1}{2} \epsilon^{\mu\nu\rho\sigma}
F_{\rho\sigma}$
and $Q_{L,R}=U_{L,R},D_{L,R}$ are vectors containing the left-handed
(LH) and right-handed (RH) quarks of the three generations (capital
letters denote matrix quantities, $q,u,d$ are used otherwise). In 
\KSVZ\ the PQ charge matrices $X_{Q_{L,R}}$ vanish, while in \DFSZ\ they
are non-zero but generation blind, hence the current in
\eqn{LabelowEW} does not dependent on the quark basis.  It is
convenient to remove the axion-gluon term via a field dependent chiral
rotation of the first generation quarks $q=u,d$:
$q_{L,R}\to e^{\mp i \frac{a}{2f_a} f_q} q_{L,R}$ with $f_u+f_d=1$.
Defining $z=m_u/m_d$ and choosing $f_u= 1/(1+z)\simeq 2/3$ avoids tree
level axion-pion mixing. As a result of this rotation the
coefficient of the QED term
gets shifted as $E/N \to E/N - f_\gamma(z)$ with
$f_\gamma \simeq 1.92$, while the axion coupling to
the first generation quarks becomes
\begin{equation}
\label{eq:firstgen}
 \mathcal{L}_{aq} = 
 \frac{\partial_\mu a}{2f_a} \sum_{q=u,d}\!
 \left[
\bar q \gamma^\mu\gamma_5 \left(\frac{X_{q_R}\!-\!X_{q_L}}{2N}-f_q\right)\! q 
\right].  
\end{equation}
The charge dependent part of the couplings is commonly denoted as
$C^0_q=(X_{q_R}-X_{q_L})/(2N)$, while the vector couplings vanish upon
integration by part because of the equation of motion.  Matching
\eqn{eq:firstgen} with the non-relativistic axion-nucleon Lagrangian
allows to extract the axion couplings to the nucleons
$N=p,n$~\cite{diCortona:2015ldu} which are defined in analogy to the
couplings to the quarks by
$\partial_\mu a/(2f_a) C_N \bar N \gamma^\mu\gamma_5 N$.  It is
convenient to recast the results in terms of the two linear
combinations
\begin{align} 
\label{CppCn}
C_p +C_n &= 
0.50(5)\;\left( C^0_u  + C^0_d  -1\right) - 2 \delta_s  \, ,  \\
C_p -C_n &= 1.273(2)\; ( C^0_u  - C^0_d  - \frac{1}{3}) ,
\label{CpmCn}
\end{align}
where the two numbers in parenthesis correspond to $f_u+f_d=1$ (exact)
and $f_u-f_d\simeq 1/3$ (approximate), while $\delta_s$ is a
correction appearing in \DFSZ\ which is dominated by the
$s$-quark sea contribution. 
In the models below, using the results
from~\cite{diCortona:2015ldu} and allowing for the 
largest possible values of $C^0_{s,c,b,t}$, we have
$|\delta_s| \lsim 0.04$.  \Eqn{CppCn} makes clear why it is
difficult to decouple the axion from the nucleons. For \KSVZ\
$C^0_u=C^0_d=0$ and the model independent contribution survives. For 
\DFSZ\ we see from \eqn{eq:firstgen} that $C^0_u+C^0_d=N_l/N$ with 
$N_l$ the contribution to the QCD anomaly of the first generation
(light) quarks. Hence, for generation blind charges 
$C^0_u+C^0_d=1/3$ is an exact result.

\Sec{The nucleophobic axion.}
We take as the defining condition for the nucleophobic axion the
(approximate) vanishing of the relations in \eqns{CppCn}{CpmCn}.
Remarkably, since the axion-pion coupling is
proportional to the isospin breaking combination $C_p-C_n$
\cite{Kim:2008hd}, nucleophobic axions are also pionphobic. 
%
We start by studying \eqn{CppCn}. In the approximation in which
$\delta_s$ is neglected, $C_p+C_n=0$ implies $C^0_u + C^0_d = N_l/N=
1$. This can only be realized in two ways: $(i)$ either the
contributions of the two heavier generations cancel each other
($N_2=-N_3$ and $N_l=N_1$) or $(ii)$ they vanish identically, in which
case it is convenient to assign $N_l=N_3$ and, hoping that no
confusion will arise with the usual generation ordering, require for
the anomalies of the heavier generations $N_1=N_2=0$.\footnote{ We
  have found that this second case was already identified in the
  not-well-known work in Ref.~\cite{Hindmarsh:1997ac}.}
\nocite{Hindmarsh:1997ac}
%
Clearly both cases require generation dependent PQ charges.  A generic
matrix of charges for a LH or RH quark $q$ can be written as
$X_Q=X^0_q I+X^8_q\lambda_8+X^3_q\lambda_3$ where $I ={\rm diag
}(1,1,1)$ is the identity in generation space, while $\lambda_8 ={\rm
  diag }(1,1,-2)$ and $\lambda_3 ={\rm diag }(1,-1,0)$ are
proportional to the corresponding $SU(3)$ matrices. In this Letter we
are mainly interested in a proof of existence for nucleophobic axions,
so we introduce some simplification: we assume just two Higgs doublets
$H_{1,2}$ (with PQ charges $X_{1,2}$ and hypercharge $Y=-1/2$), and we
consider only PQ charge assignments that do not forbid any of the SM
Yukawa operators.  Under these conditions, it can be shown that two
generations must have the same PQ charges~\cite{inpreparation}.  We
can then drop the $SU(2)$ breaking $\lambda_3$ term so that the matrix
$X_Q=X^0_q I + X^8_{q}\lambda_8$ respects a $SU(2)$
symmetry acting on the generation indices $\{1,2\}$, and we henceforth
refer to such a structure as $\mathit{2} + \mathit{1}$.
%
%
To study which Yukawa structures can enforce the condition $N=N_l$ it
is then sufficient to consider just one of the generations
in~$\mathit{2}$ together with the generation in $\mathit{1}$ carrying
index $\{3\}$. The relevant Yukawa operators read:
\begin{align}
\nonumber
& \bar q_2  u_2 H_1,\ \;   \bar q_3  u_3 H_a, \ \ \bar q_2 u_3 H_b,
   \ \ \ \  \quad    \bar q_3  u_2 H_{1+a-b}, \\
  \label{eq:11d}
& \bar q_2  d_2 \tilde H_c,\ \;    \bar q_3  d_3 \tilde H_d, \ \;
  \bar q_2 d_3  \tilde H_{d+a-b},   \ \;    \bar q_3  d_2 \tilde H_{c-a+b}, 
\end{align}
where $\tilde H =i\sigma_2 H^*$, assigning $H_1$ to the first term is
without loss of generality and, according to our assumptions, all the
Higgs sub-indices must take values in $\{1,2\}$. It is easy to verify
that in each line the charges of the first three quark-bilinears
determine the fourth one, e.g.  $X(\bar q_3 u_2) = X(\bar q_2 u_2) +
X(\bar q_3 u_3) - X(\bar q_2 u_3)$, while the third term in the second
line is obtained by equating $X_{q_3}-X_{q_2}$ as extracted from the
second and third terms of both lines.  It is now straightforward to
classify all the possibilities that yield $N_l/N=1$. Denoting the
Higgs ordering in the two lines of \eqn{eq:11d} with their indices
$\in\{1,2\}$, e.g. $(H_1,H_2,H_1,H_2)_u\sim (1212)_u$ we have
respectively for $(i_{1,2})$ $N_1=N_2=-N_3$ and $(ii_{1,2})$
$N_1=N_2=0$:
%
%
\begin{align}
\nonumber
&\!\!\!\,  (i_1)\! : (1212)_u \; (2121)_d; \ \ \ \;  (i_2)\! : (1221)_u\; (2112)_d\,;\\
\label{eq:i-ii}
& \!\!\!\! (ii_1)\! : (1111)_u \; (1221)_d; \ \ \   (ii_2)\! : (1221)_u\; (1111)_d\,.
\end{align} 
It is easy to verify that in $(i_{1,2})$
$2N_l=2N_2 = X_{u_{2R}} \!+\! X_{d_{2R}}\! -\! X_{u_{2L}}\!-\! X_{d_{2L}}= X_2-X_1$
with  $N_3 =-N_2$, in $(ii_1)$ $2N_l\!=\!2N_3=X_2-X_1$ and in
$(ii_2)$ $2N_l\!=\! 2N_3=-X_2+X_1$ with, in both last cases, $N_1=N_2 =0$.
Let us now discuss how the second condition $C_p-C_n\approx 0$ can be
realized. We denote by $\tan\beta= {v_2}/{v_1}\,, $ the ratio of the
$H_{1,2}$ VEVs, and we use henceforth the shorthand notation
$s_\beta=\sin\beta$, $c_\beta=\cos\beta$. The ratio
$X_1/X_2=-\tan^2\beta$ is fixed by the requirement that the PQ
Goldston boson is orthogonal to the Goldston eaten up by the
$Z$-boson~\cite{Dine:1981rt}, and the charge normalization is given 
in terms of the light quark anomaly as  $X_2-X_1= \pm 2 N_l$.
Here and below the upper sign holds for $(i_{1,2})$ and $(ii_1)$, and the
lower sign for $(ii_2)$.
From \eqn{eq:i-ii} it follows that in all cases
$C^0_u-C^0_d= 
-\frac{1}{2N}(X_1+X_2)=\pm(s^2_\beta-c^2_\beta)$.
The second condition for nucleophobia $C^0_u-C^0_d=1/3$ is then
realized for $s^2_\beta=2/3$ in $(i_{1,2})$ and $(ii_1)$, and for
$s^2_\beta=1/3$ in $(ii_2)$.  We learn that even under some
restrictive assumptions, there are four different ways to enforce
nucleophobia. More possibilities would become viable by allowing for
PQ charges that forbid some Yukawa operator~\cite{inpreparation}.
Note that while $C_p-C_n\approx 0$ requires a specific choice
$\tan\beta \approx \sqrt{2}$, $1/\sqrt{2}$, $C_p+C_n\approx 0$ is
enforced just by charge assignments.  For both values of $\tan\beta$
the top Yukawa coupling remains perturbative up to the Planck scale,
however, we stress that these values should be understood as relative
to the physical VEVs, rather than resulting from a tree level scalar
potential. This is because the large $v_a$ would destabilize any
lowest order result for $v_{1,2}$. This is of course a naturalness
issues common to all invisible axion models.

Finally, to render the axion invisible, $H_{1,2}$ need to be coupled
via a non Hermitian operator to the scalar singlet $\phi$ with PQ
charge $X_\phi$. This ensures that the PQ symmetry gets spontaneously
broken at the scale $v_a \gg v_{1,2}$ suppressing efficiently all
axion couplings.  There are two possibilities: $H^\dagger_2 H_1\phi$
in which case $|X_\phi| = 2N_l=2N$, the axion field has the same
periodicity than the $\theta$ term and the number of domain walls (DW)
is $N_{DW}=1$, or $H^\dagger_2 H_1\phi^2$ in which case
$|X_\phi| = N_l=N$ and $N_{DW}=2$. In contrast, in DFSZ models
$|X_\phi|=2N/3, (2N/6)$ yield $N_{DW}=3, (6)$ and a  DW
problem is always present.

\Sec{Flavor changing axion couplings.}
Generation dependent PQ charges imply FV axion couplings.
Plugging $X_Q=X^0_q I + X^8_{q}\lambda_8$
in \eqn{LabelowEW} it is readily seen that a misalignment between the
Yukawa and the PQ charge matrix becomes physical. Since we are mostly
interested in the light quark couplings, we single
out $X_{q_1}$ for case $(i)$, and $X_{q_3}$ for $(ii)$:
\begin{equation}
  \label{eq:6} 
X_Q\ =\  X_{q_1}\, I - 3 X^8_q\, \Lambda \ =\  X_{q_3}\, I + 3 X^8_q\, \Lambda'  
\end{equation}
with $3 X_q^8 = X_{q_1}\!-\! X_{q_3}$,
$\Lambda =\frac{1}{3}\left(I\!-\!\lambda_8\right) ={\rm diag}(0,0,1)$
and
$\Lambda' =\frac{1}{3}\left(2 I\!+\!\lambda_8\right) ={\rm
  diag}(1,1,0)$.
In case $(i)$, the matrices of couplings in the Yukawa basis read:
\begin{align}
\label{C0V}
& \!\!\! C^{0V}_Q= - \frac{3}{2N}\left[
 X_{q_R}^8 W_{Q_R}+ X^8_{q_L}W_{Q_L}\right], \\
\label{C0A}
& \!\!\! C^{0}_Q\!+\!\Delta C^0_Q \!=\! C^0_{q_1} I
\!-\! \frac{3}{2N} \left[X_{q_R}^8 W_{Q_R}\!-\! X^8_{q_L}W_{Q_L}\right]\!, 
\end{align}
where for $C^{0V}_Q$ the equations of motion imply  only 
the vanishing of the diagonal entries, but not of the 
off-diagonal ones, $C^0_Q = C^0_{q_1} I$ with $C^0_{q_1}$ defined
below \eqn{eq:firstgen}, and denoting by $V_Q$ the unitary rotations
to the diagonal Yukawa basis, $W_Q = V^\dagger_Q \Lambda V_Q$. While
in the models discussed here $W_{Q_R}$ and $W_{Q_L}$ are never
simultaneously present, this is possible in more general
cases~\cite{inpreparation}.  It is now convenient to single out the
diagonal (denoted by $\delta$) and off-diagonal (denoted by $\omega$)
entries in $W_Q = \delta_Q + \omega_Q$:
\begin{align}
  \label{eq:8} 
\begin{split}
&(\delta_Q)_{ij} = \delta_{q_i}\,\delta_{ij}\,, \qquad \sum\nolimits_i  \delta_{q_i} =1\,,      \\ 
&(\omega_Q)_{ii} =0\,, \qquad\qquad  |(\omega_Q)_{ij}|^2 = \delta_{q_i}\, \delta_{q_j}, \ 
\end{split}
\end{align}
where the condition on $\delta_{q}$ follows from ${\rm Tr}(W_Q)=1$,
the one on $\omega_Q$ from the vanishing of the principal minors for
the rank one matrix $W_Q$, and $\delta_{ij}$ in the first relation is
the usual Kronecker symbol. In $(ii)$ the couplings are given by
\eqs{C0V}{C0A} by replacing $C^0_{q_1}\!  \to\! C^0_{q_3}$, $(-3)\!
\to\! (+3)$ and $W_Q\! \to\! W'_Q= V^\dagger_Q \Lambda' V_Q$, while
the two conditions read $\sum_i \delta'_{q_i} =2$ and $
|(\omega'_Q)_{ij}|^2 = (1 - \delta'_{q_i}) (1 - \delta'_{q_j})$.
Information on the LH matrices can be obtained from the CKM matrix:
$V^\dagger_{U_L}
V_{D_L} = V_{CKM} \approx I$ implies $V_{U_L}\approx
V_{D_L}$ and hence $W_{U_L} \approx
W_{D_L}$. Therefore, to a good approximation we can define a single
set of LH parameters $\delta_L\!
= \!\delta_{u_L}\!\approx
\delta_{d_L}$.  In contrast, we have no information about the RH matrices
so that in general $W_{U_R}\neq
W_{D_R}$ and $\delta_{u_R},\delta_{d_R}$ are two independent sets.
Corrections to the diagonal axial couplings due to quark mixing are
listed in Table~\ref{Quarks}. Corrections to the second condition for
nucleophobia can be always compensated by changing appropriately the
value of $\tan\beta$ to maintain $C_p-C_n\approx 0$. However, this is
not so for the first condition, for which large mixing corrections
would spoil $C_p+C_n\approx 0$.  Actually, only for $(ii_1)$ a
relatively small correction can improve nucleophobia, and this is
because only in this case $C^0_s$, which determines the sign of
$\delta_s$ in~\eqn{CppCn}, is negative ($C^0_s=-s^2_\beta$), rendering
possible a tuned cancellation $-0.50\, \delta'_{d_{3R}} + 2
|\delta_s|\approx 0$, while for all other cases the value of $g_{aN}$
is increased.  Nucleophobia thus generically requires that the quark
Yukawa and the PQ charge matrices are aligned to a good approximation
(for recent attempts to connect axion physics to flavor dynamics see
e.g.~\cite{Ema:2016ops,Calibbi:2016hwq,Arias-Aragon:2017eww}).

%
\begin{table}[t]
\centering
\def\arraystretch{1.4}
\begin{tabular}{|c|c|c|c||c|c|c||c|c|c||}
\hline
&  $E_Q/N$ & $\Delta C^0_u $ & $\Delta C^0_d$
& $\left|C^{0}_{u}\right|_{i\neq j} $ & $\left|C^0_d\right|_{i \neq j} $  \\
\hline
\multicolumn{1}{|c|}{$(i_1)$}      
& $-4/3+6s^2_\beta$     &   $ - \delta_{1L}$ &   $ - \delta_{1L} $  &  $ \omega_{L}$ &  $\omega_{L}$ \\
\multicolumn{1}{|c|}{$(i_2)$}        
& $-4/3+6s^2_\beta$    & $  -\delta_{u_{1R}}$  &  $ - \delta_{d_{1R}}$  &  $ \omega_{u_R}$ &  $ \omega_{d_R}$   \\
  \hline
  \multicolumn{1}{|c|}{$(ii_1)$} & $2/3+6 s^2_\beta$   & $0$ &  $-\delta'_{d_{3R}}$ & $0$ &$\omega'_{d_R}$
  \\
  \multicolumn{1}{|c|}{$(ii_2)$} & $8/3-6 s^2_\beta$   & $-\delta'_{u_{3R}}$ & $0$  &  $\omega'_{u_R}$ & $0$\\
\hline
\end{tabular}
\caption{\label{Quarks} 
  Contributions from the quarks to $E/N$, and corrections to the
  nucleophobic  axion couplings due to quark mixings. 
  The (off-diagonal)  vector couplings  $C_q^{0V}$ are equal in
  modulus to the axial-vector ones.}
\vspace{-.3cm}
\end{table}

\Sec{Electrophobia.} 
Electrophobia can be implemented exactly (at the lowest loop order),
or approximately (modulo lepton mixing corrections) by introducing an
additional Higgs doublet 
uncharged under the PQ symmetry, and by coupling it respectively to
all the leptons, or just to the electron.  However, electrophobia can
also be implemented without enlarging the Higgs sector at the cost of
a fine tuned cancellation between $C^0_e$
and a mixing correction. Of course this requires large lepton mixings
and fine tuning. Given that large mixings do characterize the lepton
sector, at least the first requirement is not unnatural.  It is a bit
tedious but straightforward to verify that in all the following cases
a cancellation is possible:  we can assign the electron: $(i_l)$
to the doublet in $2+1$,
or $(ii_l)$
to the singlet, and in both cases we can consider $(12\,.\,.)_l$
or $(21\,.\,.)_l$
structures, and then combine these possibilities with the four quark
cases. Moreover, for $(abab)_l$
type of structures electrophobia is enforced by a cancellation from LH
mixing, while for $(abba)_l$
from RH mixing. All in all, there are $2\times
2\times 4\times
2=32$ physically different astrophobic models. However, as regards the
axion-photon coupling, there are only four different values of
$E/N$.
We have listed them in Table~\ref{Models} by picking out four
representative models.

\begin{table}[h!]
\centering
\def\arraystretch{1.4}
\begin{tabular}{|l|c|c|c|c|}
  \hline
  &  $E_L/N$ & $E/N$ &$C^0_e $ & $\Delta C^0_e$   \\
  \hline
$M1:(i)+(i_l)$ & $2-6s^2_\beta$ & $2/3$ & $ -s^2_\beta \; (-2/3)$ &   
            $ + \delta_{e_1} $    \\
  \hline
$M2:(ii_1)+(i_l)$ & $2-6s^2_\beta$ & $8/3$  & $  -s^2_\beta\; (-2/3) $ &    
            $ + \delta_{e_1}$      \\
 \hline
$M3:(ii_2)+(ii_l)$ & $-4+6s^2_\beta$ & $-4/3$  & $  s^2_\beta\; (1/3) $ &    
            $ - \delta'_{e_3}$      \\
\hline
$M4:(ii_1)+(ii_l)$ & $4-6 s^2_\beta$ & $14/3$ & $-s^2_\beta\;(-2/3)$ & $ + \delta'_{e_3}$ \\   
  \hline
\end{tabular}
\caption{\label{Models} 
  Contributions of the leptons and total values of $E/N$ in four 
  representative models, selected by the (arbitrary) choice 
that the electron couples to $\tilde H_1$. 
 The numerical values of $C^0_e$ are given in
  parenthesis, and the corrections $\Delta C^0_e$ can come from RH or 
  LH mixings. 
}
\vspace{-.3cm}
\end{table}
%
%
%
\Sec{Phenomenology of the heavy axion window.} 
We denote as $g_{af} = C_{f} m_f / f_a$ the axion coupling to
$f=p,n,e$, including corrections from mixing effects, and by
$g_{a\gamma} = \alpha/(8 \pi f_a) (E/N-f_\gamma)$ the axion coupling
to photons. The most relevant astrophysical bounds
are~\cite{Olive:2016xmw,Raffelt:2006cw}:

\noindent
$\bullet$ 
$\abs{g_{a\gamma}} < 6.6 \cdot 10^{-11}$ GeV$^{-1}$ ($95 \%$ CL)  from the evolution of HB 
stars in globular clusters \cite{Ayala:2014pea}.

\noindent
$\bullet$ 
$\abs{g_{ae}} < 2.7 \cdot 10^{-13}$  $(< 4.3 \cdot 10^{-13})$  
($95 \%$ CL) from the shape of the WD luminosity function
\cite{Bertolami:2014wua}  
(from RG evolution in globular clusters \cite{Viaux:2013lha}).

\noindent
$\bullet$ $g_{ap}^2 + g_{an}^2 < 3.6 \cdot 10^{-19}$ from the SN1987A
neutrino burst duration~\cite{Giannotti:2017hny}.  Large uncertainties
in estimating SN axion emissivity~\cite{Keil:1996ju,Fischer:2016cyd}
prevent assigning a reliable statistical significance to this limit.

\noindent
$\bullet$ Structure-formation arguments also provide hot DM (HDM)
limits on the axion mass: in benchmark
models $m_a \lesssim 0.8\,$eV~\cite{Archidiacono:2013cha,DiValentino:2015wba,Olive:2016xmw}.
However, nucleophobic axions are also pionphobic, and the main 
thermalization process $\pi \pi \to \pi a$ is then suppressed, relaxing the HDM
bound. This also implies that large-volume surveys like EUCLID
\cite{Laureijs:2011gra} cannot probe astrophobic axions.
%

\begin{figure}[t!]
\includegraphics[width=.5\textwidth]{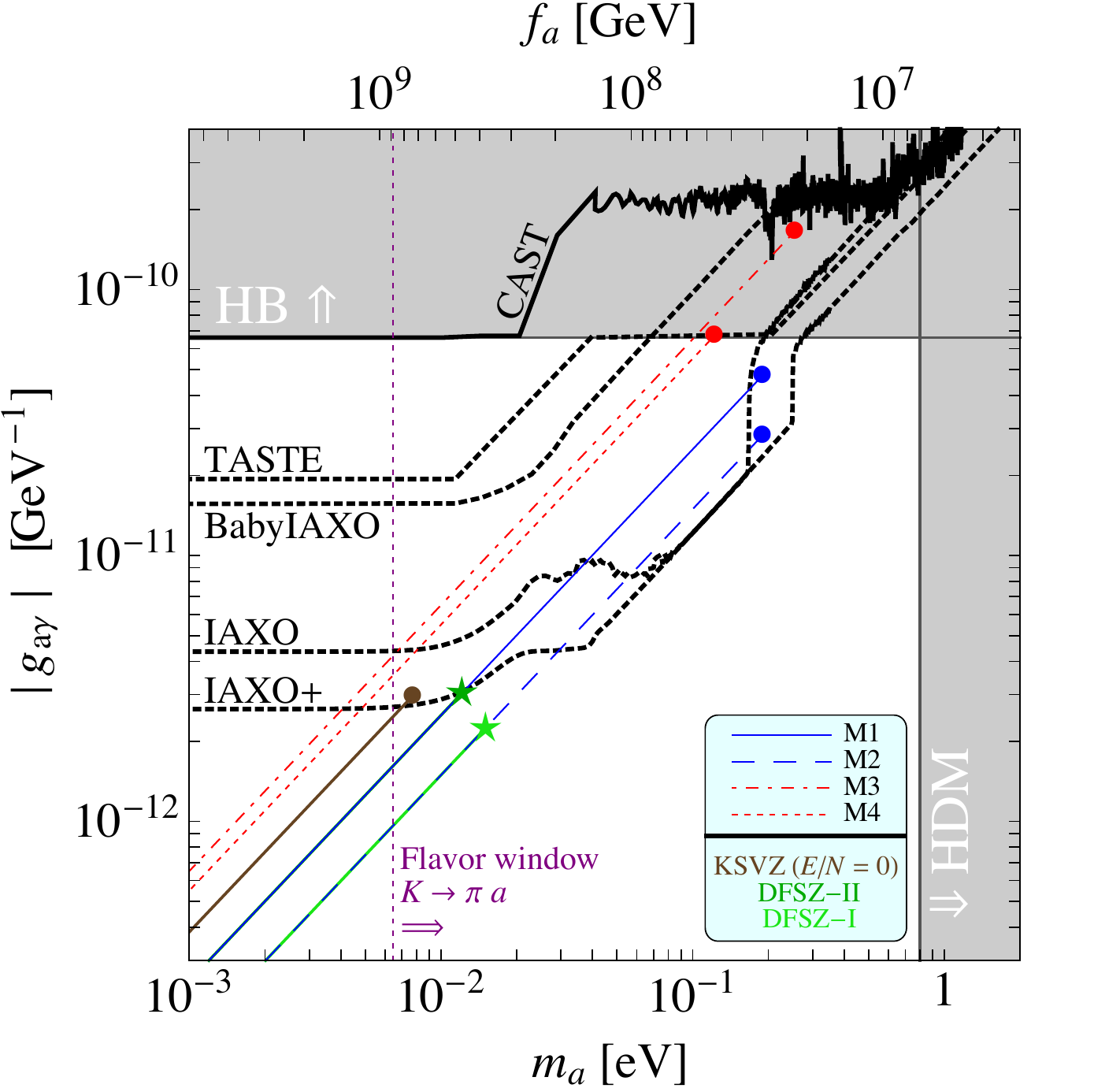}
\caption{\label{fig:Excl}Axion-photon coupling $|g_{a\gamma}|$ for the
  astrophobic models in Tab.~\ref{Models} as a function of $m_a$.  The
  \DFSZI-II\ (respectively with $E/N=8/3,2/3$) and \KSVZ\ benchmarks are
  also shown for comparison.}
\end{figure}

The main results for astrophobic axions are summarized
in~\fig{fig:Excl} and compared to the \KSVZ\ and \DFSZ\
benchmarks. The lines are broken at: {\large $\bullet$}~marks, which
indicate the upper bounds on $m_a$ from SN1987A, and {\large $\star$}
marks, corresponding to the combined SN/WD constraints for
\DFSZ\ models.
%
%
As anticipated, for \KSVZ\ and \DFSZ,  axion masses above
$m_a \sim 10^{-2}$ eV are precluded by the SN/WD limits (dark brown
bullet for \KSVZ\ and green stars for \DFSZ). For
astrophobic axions the SN/WD bounds get significantly relaxed 
(they cannot evaporate completely because of the contribution 
$\delta_s$ in Eq.~(\ref{CppCn})
to $g_{aN}$).
We obtain $m_a < 0.20\,$eV for \MI/\MII\ (blue
bullets), $m_a < 0.25\,$eV for \MIII\ and $m_a < 0.12\,$eV for \MIV\
(red bullets).

\smallskip $\diamond$ \emph{Searches with helioscopes:} Helioscopes
are sensitive to $g_{a\gamma}$ which is not particularly suppressed in
astrophobic models. The solid black line in Fig.~\ref{fig:Excl} shows
the present limits from CAST~\cite{Anastassopoulos:2017ftl}, while the
dotted black lines show the projected sensitivities of next
generation helioscopes. While the improvement in mass reach will be
limited for TASTE~\cite{Anastassopoulos:2017kag} and
BabyIAXO~\cite{Irastorza:2017}, we see that IAXO
\cite{Armengaud:2014gea,Irastorza:1567109} and its upgrade IAXO+
\cite{Giannotti:2017hny} will be able to cover the whole interesting
region up to $m_a\sim 0.2$ eV.

\smallskip $\diamond$ \emph{Flavor Violation Experiments:} The
strongest limits on FV axion couplings to quarks come from
$K^+ \to \pi^+ a$~\cite{Hindmarsh:1998ph}. Comparing the 
model prediction with the current limit~\cite{Adler:2008zza} gives
\begin{equation}
\label{Kpia}
{\mathcal B}_{K^+ \to \pi^+ a} \simeq 10^{-2}\! 
\left( \frac{m_a}{0.2 \eV} \right)^2\! \omega^2_{l2} \lesssim 7.3\cdot 10^{-11} , 
\end{equation}
where
$\omega^2_{l2}=|\omega_{12}|^2=\delta_{1L}\delta_{2L},\,\,\delta_{d_{1R}}\delta_{d_{2R}}$
for $(i_{1,2})$, $\omega^2_{l2}=|\omega'_{32}|^2$ for $(ii_1)$, while
in $(ii_2)$ the branching ratio vanishes (see Table~\ref{Quarks}).
For models $(i_{1,2})$, by taking CKM-like entries in $V_{d_{L,R}}$,
$\omega_{12}^2 \gsim 10^{-8}$ and the limit would be saturated.
This implies that NA62, which is expected to improve by a factor of
$\sim$70 the limit on $K^+ \to \pi^+
a$~\cite{Anelli:2005ju,Fantechi:2014hqa} can probe these models (the
explorable ``flavor window" corresponds to the vertical magenta line
in Fig.~\ref{fig:Excl}).  In contrast, in case $(ii_1)$ for CKM-like
mixings the limit \eqn{Kpia} would be exceeded by six orders of
magnitude,
which renders this case rather unrealistic.  If we allow for only two
Higgs doublets, electrophobic models necessarily have FV axion
couplings to leptons.  The strongest limits come from searches for
$\mu\to e\gamma a$~\cite{Bolton:1988af, Goldman:1987hy}. They yield
$f_a/\omega_{e\mu} \gsim 2 \cdot 10^9\,$GeV which implies
$m_a \lesssim 2.7\cdot 10^{-3}/\omega_{e\mu}\eV$.  Recalling that
$\omega_{e\mu}=\sqrt{\delta_e\delta_\mu}$ and that
$\delta_e \sim O(1)$ is needed to cancel $C^0_e$, we need to impose
$\delta_\mu \lesssim 10^{-4}$ to avoid overconstraining the large mass
window. This implies $\delta_\tau \sim O(1)$ from which we can predict
%
${\mathcal B}_{\tau \to e a} \simeq 7 \cdot 10^{-6} \left( \frac{m_a}{0.2 \eV} \right)^2 \, ,$ 
about three orders of magnitude below the present bound~\cite{Albrecht:1995ht}.
  
\smallskip $\diamond$ \emph{Axion DM in the heavy mass window:} For
$m_a \sim 0.2$ eV the misalignment mechanism cannot fulfill
$ \Omega_a \simeq \Omega_{\rm DM}$.  In post-inflationary scenarios,
if $N_{\rm DW} > 1$~\cite{Kawasaki:2014sqa,Ringwald:2015dsf} an
additional contribution from topological defects can concur to
saturate $ \Omega_{\rm DM}$. This requires an explicit PQ breaking to
trigger DW decays, and a fine tuning not to spoil the solution to the strong CP problem.
For $N_{\rm DW} = 1$ a contribution to the relic abundance
can come from axion production via a parametric resonance in the
oscillations of the axion field radial mode~\cite{Co:2017mop}, in
which case  $f_a \lsim 10^{18}\,$GeV is needed.  In both
cases, the lower values of $f_a$ allowed by the astrophobic models can
help to match the required conditions.

{$\diamond$ \emph{Stellar cooling anomalies:}} Hints of anomalous
energy loss in stars \cite{Giannotti:2015kwo,Giannotti:2017hny} can be
more easily accommodated in astrophobic axion models.
Ref.~\cite{Giannotti:2017hny} finds, as the best-fit point for extra
axion cooling, $g_{a\gamma}\sim 0.14 \cdot 10^{-10}$ GeV$^{-1}$ and
$ g_{ae} \sim 1.5 \cdot 10^{-13}$.  While in the \DFSZ\ model this
point is in tension with the SN bound, it is comfortably within the
allowed parameter space of the astrophobic axion.

\Sec{Conclusions.} We have discussed a class of \DFSZ-like axion
models with generation dependent PQ charges that allows to relax the
SN1987A bound on $g_{aN}$ and the WD/RG limit on $g_{ae}$,  and to 
extend the viable axion mass window up to $m_a\sim 0.2\,$eV.
This scenario is characterized by compelling connections with flavor
physics.  Complementary informations to direct axion searches can be
provided by experimental searches for FV meson/lepton decays and,
conversely, the discovery of this type of astrophobic axions would
provide evidences that the quark Yukawa matrices are approximately
diagonal in the interaction basis, conveying valuable information on
the SM flavor structure. While we have restricted our analysis to PQ
charge assignments which do not forbid any of the SM Yukawa operators,
it would be interesting to relax this condition, and explore to which
extent the PQ symmetry could play a role as a flavor symmetry in
determining specific textures for the SM Yukawa matrices.

\section*{Acknowledgments}
We thank Michele Redi for helpful discussions and collaboration at the
early stages of this project, Maurizio Giannotti for communications
regarding Ref.~\cite{Giannotti:2017hny}, Torben Ferber and
Massimiliano Lattanzi for useful comments.  L.D.L, F.M., P.P. and
R.Z. acknowledge hospitality and financial support from the LNF Theory
group where this work was started, E.N and L.D.L.  acknowledge the
CERN Theory group for hospitality and financial support during the
development of the project.
F.M.~is supported by MINECO grant FPA2016-76005-C2-1-P and by Maria de Maetzu
program grant MDM-2014-0367 of ICCUB and 2014-SGR-104. E.N.~is
supported in part by the INFN ``Iniziativa Specifica'' TAsP-LNF.

\bibliographystyle{apsrev4-1.bst}

%

\end{document}